# Deep Learning for Breast Cancer Classification: Enhanced Tangent Function

Ashu Thapa[1], Abeer Alsadoon[1,2,3,4,*], P.W.C. Prasad[1,5], Simi Bajaj[2], Omar Hisham Alsadoon[6]
Tarik A. Rashid[7], Rasha S. Ali [8], Oday D. Jerew[4]

[1]School of Computing and Mathematics, Charles Sturt University (CSU), Wagga Wagga, Australia
[2]School of Computer Data and Mathematical Sciences, University of Western Sydney (UWS), Sydney, Australia
[3]Kent Institute Australia, Sydney, Australia
[4]Asia Pacific International College (APIC), Sydney, Australia
[5]Australian Institute of Higher Education, Sydney, Australia
[6]Department of Islamic Sciences, Al Iraqia University, Baghdad, Iraq
[7]Computer Science and Engineering, University of Kurdistan Hewler, Erbil, KRG, IRAQ
[8]Department of Computer Techniques Engineering, AL Nisour University College, Baghdad, Iraq

Abeer Alsadoon[1*]
* Corresponding author. A/Prof (Dr) Abeer Alsadoon, [1]School of Computing and Mathematics, Charles Sturt University, Wagga Wagga, Australia. Email: alsadoon.abeer@gmail.com , Phone +61 413971627

## Abstract

*Background and Aim:* Recently, deep learning using convolutional neural network has been used successfully to classify the images of breast cells accurately. However, the accuracy of manual classification of those histopathological images is comparatively low. This research aims to increase the accuracy of the classification of breast cancer images by utilizing a Patch-Based Classifier (PBC) along with deep learning architecture. *Methodology:* The proposed system consists of a Deep Convolutional Neural Network (**DCNN**) that helps in enhancing and increasing the accuracy of the classification process. This is done by the use of the Patch-based Classifier (PBC). CNN has completely different layers where images are first fed through convolutional layers using hyperbolic tangent function together with the max-pooling layer, drop out layers, and SoftMax function for classification. Further, the output obtained is fed to a patch-based classifier that consists of patch-wise classification output followed by majority voting. *Results:* The results are obtained throughout the classification stage for breast cancer images that are collected from breast-histology datasets. The proposed solution improves the accuracy of classification whether or not the images had normal, benign, in-situ, or invasive carcinoma from 87% to 94% with a decrease in processing time from 0.45 s to 0.2s on average. *Conclusion:* The proposed solution focused on increasing the accuracy of classifying cancer in the breast by enhancing the image contrast and reducing the vanishing gradient. Finally, this solution for the implementation of the Contrast Limited Adaptive Histogram Equalization (CLAHE) technique and modified tangent function helps in increasing the accuracy.

## Keywords

## 1. Introduction

Breast cancer is the most commonly occurring invasive cancer in women. It is the second main cause of death in women after the lung cancer [1]. The identification of cancer depended on the digital biomedical photography analysis (histopathological or biopsy images). Contradictory decisions are produced by the doctor and radiologists depending on the investigation of such images. Hence, it is prone to errors. Screening mammography is a common traditional technique used to read and analyze mammogram images. It was evaluated by human readers. This technique is tiring, monotonous, costly, importantly prone to error and lengthy process[1]. Since the rate of women suffering/dying from breast cancer has been increasing, there is a need for identifying and detecting the mass lesion at its early stage. So, in this paper we propose an alternative method to recognize whether the breast has a benign or malignant type of tumor depending on a computer-based image-processing and machine learning system [2]. This system utilises a deep learning technique, convolutional neural network (CNN) for the classification and diagnosis of breast cancer. Convolutional and pooling layers of CNN help in handling the high dimension of images. The focus is a system that can provide high accuracy in classifying the type of tumor in the breast. To achieve the best performance and result we are taking a patch-based classifier which has two models. The two models are one-patch in one decision and all patches in other decision. Henceforth, convolutional neural network along with patch-based classifier (PBC) is best known for detecting, localizing, and classifying the images, and it has shown an improved rate in classifying histopathological images accuracy. This ultimately reduces the cost, error, and workload of doctors and radiologists by increasing the accuracy rate of classification of mass lesions [3].

The development of computer vision and machine learning algorithm provides a reliable classification method with a high accuracy rate. Deep learning has been a modern method that is utilized to identify and classify the patterns of images. Among different techniques of deep learning, convolutional neural network is widely used and accepted. It provides the benefit of not only classifying the cancerous image but also extracts its important features [4]. CNN consists of multiple layers namely, convolutional layer, max-pooling layer, fully connected layer, and a drop-out layer. In [3], it is utilized in learning and extracting the features of histopathological images of the breast dataset and classifying it accurately to obtain the results as cancerous breast or not. The algorithm has got some important areas to be taken care of such as the number of datasets used for training the classifier, filter size, kernel size, activation function, number of feature maps, and regularization. Henceforth, CNN has some limitations that may lead to not providing optimum accuracy results. So, it must improve its selection factor to perceive the expected outcome with a better accuracy rate and cutting down the processing time [3].

In the recent studies of a deep convolutional neural network, different techniques and algorithms have been used for learning the feature, extracting and classifying to produce a system that has the best accuracy. The result of chosen state of the art method achieved a classification accuracy of rate 77.4% first in one patch in one decision mode, and after that with all patches in one decision mode, the accuracy increased to 87% [3]. However, the classification accuracy of an image is impacted by the factors like filter size, kernel size, and the number of datasets, the activation function, and the number of the feature map. The problem behind this state of the art is the activation function which stops to acquire a better accuracy [1]. Use of hyperbolic tangent activation function causes the gradient of loss function approach to 0 which made network training to be difficult. The gradient error tends to be high when tangent activation function is used. With this problem, the accuracy of the classification degrades automatically. Hence, current studies can still be improved.

This paper proposes a system for breast cancer detection as an enhancement to state of the art system. This system improves and enhances the classification accuracy by utilizing a patch-based classifier along with a convolutional neural network. The state of the art system does not address the problem created by the degraded contrast and noise produced by the images in terms of accuracy. This study proposes a modified hyperbolic tangent activation function which is a combination of activations on the convolutional layers as rectified linear units (ReLU) and arcTan function [1]. The proposed system solves the problem of vanishing gradient error while back-propagating by minimizing the negative points and causes a small derivative. Similarly, to enhance the contrast of the image, CLAHE method is used [5]. The proposed system can boost the degraded contrast of the images by transforming pixel intensity to a value within a display range appropriate to a pixel intensity's level in a local intensity histogram. Here, the images are enhanced by utilizing a clip level that reduces edges shadow and also noise produced in the histopathological image of breast cancer. Therefore, the overall accuracy of the system is improved with lower processing time. The comparison showed that the accuracy rate of the proposed system is higher than the state of the art.

## 2. Literature Review

A literature review is conducted to perform a survey of currently available research papers. This suvey will help in identifying existing problems and the suggested techniques that can be used in improving the prevalent problem in the state-of-the-art system being considered as base model for this research. In this section, different technologies, techniques, tools, and methods that were applied to the research and study to describe and overcome the existing problems that are known along with their level of achieved accuracy is presented. Moreover, this section includes a review for different authors on the related area of classification of breast cancer.

An enhanced k-means algorithm [1] is used to improve the accuracy and sensitivity problem in breast cancer image classification. A deep neural network guided with cell nuclei position and orientation is proposed to improve the performance. To obtain the more clear views of images, large datasets are trained by the deep neural network model so that the data augmentation and transfer learning with fine local tuning methods can be applied. They provided a convolutional neural network along with a mean-shift algorithm to overcome the problem of precision, accuracy, and overfitting. When combining CNN with the mean-shift algorithm and SoftMax layer, an accuracy of 91% and f-measure of 93% is achieved. The best accuracy, specificity, sensitivity, recall, and f-measure are 91%, 96%, 93%, 96%, and 93% respectively. To improve the model's performance with reference and accuracy the parallel feeding of local data along with raw pixels can be used [6].

A convolutional neural network is used [7], which enhanced the provided the infrared thermal image technique that is utilized to make a thermal comparison effectively between the healthy and cancerous breasts. It uses the Inception V3 model along with super machine vector to bring better accuracy by 0.78 in classifying sick or healthy breasts. The accuracy level for getting better classificationcan can be improved by increasing the pixel quality of the thermal sensitivity camera, this will help in generating a precise diagnosis [8]. The images from the large public dataset such as DDSM, BCDR, INbreast [4] being studied to classify them by utilizing different classification techniques and algorithms, i.e. VGG16, ResNet50, Inception v3 along with fine-tuning strategy and the breast cancer screening framework classifier. It provides an accuracy of 97.35% for DDSM, 95.50% for INbreast, and 96.67% for BCDR for a correct diagnosis, respectively. It gives the accuracy rate for different datasets compared to using different algorithms as a basis to compare the accuracy and performance as used in different research. For a better accurate result, the dense breast should also be considered [9].

Several multiple instance learning methods (MIL) are improved in [2], and it also reports the best method among them by showing the accuracy and limitations of each model. They offer multiple instance learning convolutional neural network (MILCNN) to solve the data augmentation problem [10] that comprises when the images are split for data augmentation in object detection; the labels are not kept. It takes data augmentation generated images as a bag by combining convolutional neural network with a specific MIL loss function derived from the bag. It was provided as an effective solution which is used to help experts in decision-making to classify images that show the level of the p-value for each method. It provides the p-values inferior to 0.001 for the MILCNN and non-parametric method which outperforms the other method which has a p-value of 0.5 and 0.0029. The authors in [11] designed a system that helps in decomposing the images into patches. The patch-based approach allows the processing of non-rectangular regions in the image by masking certain areas, by simply excluding patches from the collection. The presented convolutional neural network improvement for breast Cancer classification (CNNI-BCC) utilized the patch feature relevance to detect and classify the lesion region and category. It gives the accuracy rate for different datasets using different techniques, which can be a basis to compare the accuracy and performance of other techniques used in different research. The authors can focus on data augmentation for solving the problem of overfitting [12].

A model called You Only Learn Once was proposed in [13]. The model used the convolutional layers that are being followed by a fully connected neural network. It detects the position of mass and to identify the tumor type either benign or malignant. It is provided a feasible result along with a promising result in terms of accuracy, sensitivity, confidence factor, etc. It also helps in detecting masses that exist over the pectoral muscle and is surrounded by dense tissue, which is the most challenging case of breast cancer. When augmented datasets are used, their area under curve (AUC) reached 92.9% with an accuracy of 96.7%. The accuracy can be increased by 2% when batch normalization is applied [14] and can be tested in this model in practice for its real validity.

The work in [16] has enhanced correct classification rate with the proposed model multi-category classification of breast histopathological images using deep residual network (MuDeRN). This model utilized a stain normalization approach for improving the dimension of an image, eliminates the text annotation and black border from the image, and this is important to achieve a proper result. The feature and data of residual network (Resnet) can be trained to achieve higher accuracy in the correct classification rate (CCR) [17]. Faster R-CNN technique [8] has been adopted for malignant or benign lesions detection and classification in a mammogram without any human intervention. It has a region proposal network that helps in localizing objects to images. The model detects the malignant lesions with a sense of 0.9 and the receiver operating a curve is 0.95. The method utilized in this solution is not only bound for classification. It assists in detecting, localizing, and then classifying the tumor-based images [18]. The area under a curve obtained is 0.95 which is competent. In this solution, a small size of publicly available pixel-level annotated dataset has been utilized for training, but the classification performance of this model has been evaluated in on large screening dataset. Detection performance can only be evaluated on a small INbreast dataset.

A fine-tuned pre-trained deep neural network architecture ResNet and Inception had been applied in [19]. The data augmentation and advanced pre-processing are used to overcome the problem of over-fitting of the data [20]. The accuracy of the classification was shown according to the type and sub-types. The obtained accuracy is about 98.7% in the classification of benign and malignant breast cancer. Similarly, the obtained accuracy according to sub-type is 94.8% and 96.4%. Since the accuracy rate and utilization show sensitivities for large datasets, this can provide more convincing results. Different deep learning semantic segmentation algorithms such as U-net and DeepLab v3 can be used in the future.

A classification technique [5] is improved for large sets of data of mammogram images by utilizing CNN. They enhanced the classic approach such as using DSIFT features and SVM classifiers for two-class classification or a three-class with a CNN-DW and CNN-CT method. The error and accuracy rate are improved in comparison to [21]. They have used the augmented data to overcome the over-fitting problem. The image dataset is filtered by utilizing a CLAHE to enhance the contrast of the images. They used 2D-DWT to decompose enhanced mammogram images as its four subbands. The DSIFT descriptor has been used in this solution to extract features for all the subbands. The input data matrix is created consisting of these subband features of all mammogram patches. It is processed as an input to CNN. The SVM and softmax layer is used to train CNN for the classification process. It provides an accuracy of 81.83%. A comparison of this method with another method showed that the this solution shows a satisfactory result in classification. It has included augmented data. It enhanced them by utilizing the CLAHE method to improve the contrast of an image. It has also used DSIFT to extract further features.

The classification for both binary and multi-class histopathological images [22] had taken into consideration. It uses Inception_ResNet_V2 for the analysis of histopathological images. The better experiment was performed in the augmented datasets while doing multi-grouping, this is valuable for clinical stages. The accuracy value is 95.3 when using the Inception_ResNet_V2 model. This solution gives an acceptable range of accuracy of 95.3%, which is suitable in contrast to the earlier method [23]. This is useful for a reliable binary and multi-classification of the breast cancer images. Future work represents a reduction and avoiding the influences that affect the analysis of histopathological images of breast cancer. The performance of the classifier [24] was enhanced by combining DCNN with gradient boosting trees classifier. Inception v2 is incorporated in the methodology that demonstrates good performance at a low computational cost. Inception-300*300+GBT gives 93.5%, 95.3%, 96.1% and 91.1% accuracies on 40X, 100X, 200X and 400X BreakHis databases. In this study, the highest accuracies in detecting one or two breast cancer types can be achieved depending on using of the Inception network. However, other inception networks could be better at recognizing the remaining types. Loss function can be used to minimize intra-class distances of image patches [25]. Thus, this combination is the better solution to use the advantages of multi-resolution images and multi-scale feature descriptors and extracting both global and local information of different breast cancer tumors.

The work in [26] provided a solution to the problem of classifying breast density. Breast density categorizing is important to make an accurate classification. They utilized the ALexNet method and performed an additional analysis by removing noisy images. Noisy images may be represented as a less reliable and inaccurately labeled case in the actual clinical practice when determining two BI-RADS density categories. This leads to a boost in the classification performance with a higher area under curve i.e. 0.98, which is ultimately increased the accuracy of the solution. Here, the accuracy improvement is shown from training the CNN model from scratch to increase the training samples. It has used a fine-tuning process to increase accuracy to 0.926 which is far better than the state of the art solution [27]. Likewise, after removing the inaccurately labeled images AUC was increased to 0.9882 and 0.9857. The improvements in the identification types of images are more likely to be misclassified compared with other CNN models to AlexNet, developing strategies for dealing with noisy and inaccurate data and testing using large multi-center datasets are the future approaches.

## 2.1 State of the Art System Description

As introduced in section 1, a patch-based classifier has been introduced in [3], which is comprising of two models namely one patch in one decision and allpatches in one decision model for the classification purpose. Here, the process of coloring histopathological images and normalizing those images are held to be essential steps to decrease the variation of image appearance. This technique has been used where different scanners are used in different locations, where scanning methods may be different. The accuracy of 87% is obtained in the test of hiding data, which is considered being useful and can be improved by increasing the contrast of input images [15].

This section illustrates the features of a state of the art system by [3] where the authors proposed a patch-based classifier (PBC) by using a convolutional neural network for automatic classification of histopathological images of the breast to increase the accuracy. . Here, the process of coloring histopathological images and normalizing those images are held to be essential steps to decrease the variation of image appearance. This technique has been used where different scanners are used in different locations, where scanning methods may be different. The use of patch extraction and data augmentation helped in boosting several training samples, the research was conducted by classifying images into two different modes namely one patch in one decision (OPOD) and all patches in one decision (APOD). The patch-based classifier first predicts the class label of each patch by OPOD, if the class label is same for all the extracted patches and that is the class label of that image, then the output is supposed to be as a correct classification. This is then passed to APOD mode. Here, the class label of each extracted patch undergoes a majority voting scheme to take a final decision about the class label of the image [3]. This solution provided an accuracy rate of 77.4% first in OPOD mode which is improved by passing to APOD mode, the accuracy increases to 87 %. Each model of the state of the art system consists of multiple stages namely, image pre-processing, patch extraction and data augmentation, feature extraction using CNN and classification [3] described briefly as below.

*Pre-processing*: Stains, i.e. hematoxylin and eosin (H&E) are used to enhance the contrast between different structures (histopathological). When using this stain, hematoxylin colors the nuclei with a bluish shade, and eosin provides reddish pink. But H&E stain images are prone to color variation. This resulted in a problem to identify image nuclei correctly. So, a color normalization technique is used, which reduces the color variations. The colors in the input images were transformed into optical density with logarithmic function to normalize their concentration.

*Patch extraction and Data Augmentation*: Patch's extraction and augmentation are necessary to increase several training samples per class. Nuclei structure is necessary for analyzing breast images. Almost all nucleuses are roughly formed in a circle with a constant radius. So, from each image, the squared size patches have been extracted. Non- overlapping patches have been used to avoid redundancy of the nuclei information for one or more pixels. Image extraction is used to extract image patches with squared size, the augmented image resulting after applying the transformation operations. Due to the complexity of this model which includes numerous numbers of parameters, the chance of overfitting to the training data was increased. Data augmentation has been used to avoid such a problem[3]. It generates new sample images by applying transformations like rotation, flipping, etc to the actual sample. The combination of 90,180 and 270 degrees of rotation and flipping transformation is performed to produce seven times more images than the original database number.

*Feature extraction using Convolutional Neural network and classification:* CNN has been used by Roy et. al. in [3] to learn features from the data set matrix̃ $M$ that is obtained from the feature extraction process. CNN has a multilayered architecture which is comprising a convolution layer followed by a maximum pooling layer. The number of layers depends on the designer. The output of the final maximum pooling layer is fed to a fully connected layer that performs like a multi-layered perceptron (MLP); this is further forwarded to Softmax. It is used activation functions such as a tangent hyperbolic function. Furthermore, it is used as a pooling layer to reduce dimensions in the convolutional layer. The most used algorithms of a pooling layer are average pooling, mean pooling, and maximum pooling. During the training, the dropout layer is implemented by randomly damaging the neurons. The final layer of CNN is a softmax layer which comprises the output neuron according to the number of the problem classes, which is assigned a confidence score. The six convolutional and 5 - max-pooling layers are used as a kernel size of $3 \times 3$. The generic feature has been extracted in the initial convolution layer from input patches so that a smaller number of filters are used in the initial layer. As the succeeding layers are responsible for determining the most complex pattern among the extracted features from the previous layers, hence a greater number of filters are used in these layers. Then, three fully connected neural layer is used followed by drop out layer. Two drop out layers is used which, is preventing the over-fitting problem with probability p-0.2. The performance of the classifier is enhanced. Finally, the output layer comprising 4 neurons for classifying the labels as 0,1,2, and 3 correspondence to normal, benign, in-situ, and invasive cancer [3]. Additionally, patch-based classifier is used to predict the class label of each patch by OPOD. If the class label is the same for all the extracted patches and that is a class label of that image, then the output is supposed to correct classification. This is then passed to APOD mode. Here, the class labels of each extracted patch undergo a majority voting scheme to take a final decision about the class label of the image [3]. See table 1

**Table 1: Convolutional and max-pooling layer or method**

| Algorithm1: Convolutional method and max-pooling method of convolutional neural network for reducing input dimensions. |
|---|
| Input: Stained Image of 512*512 pixels of breast |
| Output: Feature map without the reduction of input dimension |
| BEGIN |
|     Step1: for each image of 512*512 pixel |
|     Step 2: For [iteration <0], go to Step 6 |
|     Else |
|     For [iteration <1], go to Step 2 |
|     else |
|     Step 3: set a value of respective accumulator to zero. |
|     Step 4: Input is multiplied with a respective weight as w0*x0+w2*x1+……..+wn*xn. This is then stored on the accumulator and pass forward to the next layer. |
|     Step 5: Next iteration (increasing the value of 1 by 1, i.e. the next neuron.) |
|     Next Image (continue to go to step 1) |
|     Step 6: Each perceptron consists of a count of votes. it allows us to perform a majority voting to classify the cells as per label. |
|     Step 7: Apply the hyperbolic tangent function to obtain energy output |
| END |

This model presents a classification accuracy between 77.4% to 87% when using the two modes of classification namely one patch one decision and all patch in one decision respectively, for more information about the augmented patch dataset color normalization equations see reference [3].

The six convolutions and five max-pooling filters are used to segment the image. It reduces the dimension of input to avoid the over-fitting problem. However, accuracy can be increased by solving the over-fitting problem. The neuron is the fundamental building block of the convolutional architecture. It comprises multiple neurons with learnable weights. It is also responsible for extracting discriminating features from the image and generate an output feature map.

## 3. Proposed System

After studying several methods for breast cancer image classification, we analyzed the possible pros and cons of each method. Based on the review, accuracy, activation function, overfitting problem, image enhancement, and noise hindrance, an enhancement to an existing model is proposed. From our reviewed list of the classification methods, we selected the one by Roy et. al. explained in section 2.1 [3] as the basis for our proposed solution. The reason behind the selection of this base system is that it has used a convolutional neural network for the feature extraction process and a patch-based classifier (PBC) for image classification. It leverages feature selection on several layers of CNN namely a convolutional layer, max-pooling layer, activation function, drop-out layer, SoftMax layer, followed by patch-wise classification, majority voting, and image-wise classification. The convolution layer extracts the generic features from an input patch, then passed to the following layers.

Following problems have been identified with the base system. This system causes the vanishing gradient error which is resulting in reduced accuracy of the classification. It has used a hyperbolic tangent function which is including small derivatives. Similarly, because of the image contrast and noise production, the proposed solution can read the input image accurately. Weights are updated through backpropagation to minimize loss and increase accuracy. While updating the weight, the gradient of the loss function is calculated. The activation function tanh is used here. This causes the gradient of the loss function approach to 0 making the network hard to train. The gradient error is prone to be high when the tangent activation function is used. This is called the vanishing gradient problem. When this problem arises, the accuracy of classification automatically degrades. So, in the proposed solution, this function is replaced by another modified activation function. Secondly, while the images are taken, the noise hindrance is not taken into considerations. Due to the noise, the image quality and adequate information needed to be collected, and this could not be made efficiently, which is resulted in poor classification. Therefore, image enhancement should be performed by reducing the noise problem.

This is done in the proposed solution by utilising information from Jadoon et. al. about image enhancement by filtering them through contrast limited adaptive histogram equalization (CLAHE) that helps in enhancing the degraded contrast of the mammogram images[5]. It reduces the shadow of edges along with the produced noise in the images. CLAHE is the enhanced feature of adaptive histogram equalization (AHE) which solves both the problem of noise and shadow edging.

The block diagram of the proposed solution showing enhancement of the block diagram from state of the art base system in [3] is shown in figure 1 and 2.

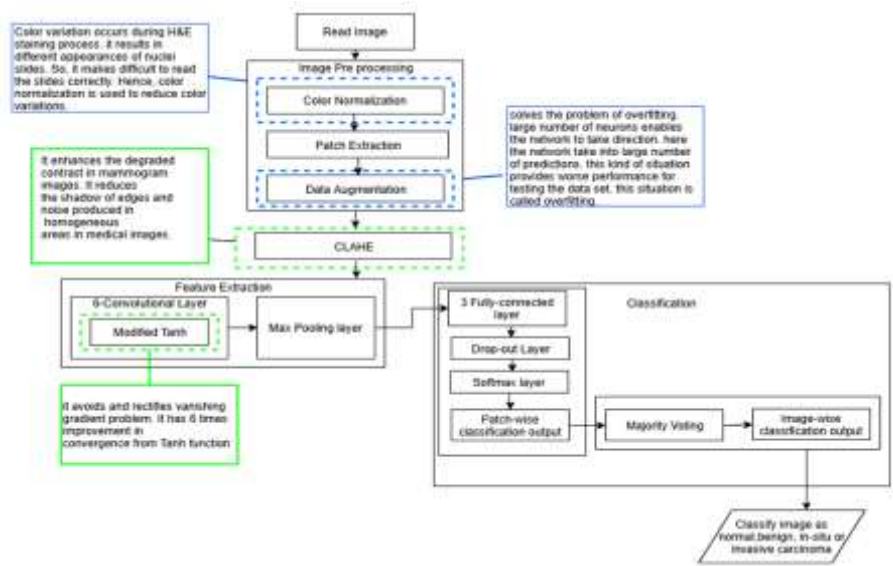

**Figure 1:** Block Diagram of the proposed deep learning system for breast cancer
(The green border refers to the new feature of the proposed system)

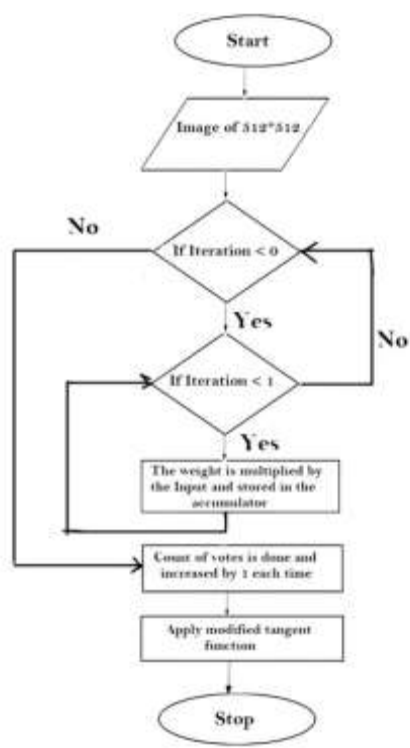

Figure 2 Enhanced Algorithm Flowchart

The proposed enhanced model has three stages namely image pre-processing, feature extraction and classification similar to [3], where contribution from this research is made in stages of pre-processing and feature extraction as shown in figure 1 with green borders. The updated stages are explained as below:

*Pre-processing:* In this stage color normalisation, patch extraction and augmentation are same as in [3]. Further, contrast limited adaptive histogram equalization (CLAHE) is used as shown in figure 1 to filter the dataset for image contrast enhancement. Through this method, the images are enhanced by the user-defined clip level. Here, the enhancement is performed on very small patches because of the over enhancement due to noise, and the effect of edge shadowing is reduced. .

***Feature extraction using convolutional neural network:*** CNN is used by [3] to learn features from the data set matrix̃ *M* that is obtained from the feature extraction process. CNN has a multilayered architecture, consisting of a convolution layer followed by a maximum pooling layer. The output of the final maximum pooling layer is fed to a fully connected layer that works like multi-layered perceptron (MLP) which is further forwarded to softmax. The original state of the art model used the activation function as tangent hyperbolic function which presented a problem of vanishing gradient error as highlighted earlier in this paper. The proposed solution uses a modified form of the tangent function, which is a combination of ReLU and ArcTan function as shown in figure 1. ReLU solves the vanishing gradient problem and leads to a small derivative whereas ArcTan is reduced by minimizing the negative points. It utilizes a pooling layer for dimensionality reduction in the convolutional layer as in state of the art model.

Then, classification is performed as in base system.

We proposed two equations. Firstly, stained images are enhanced by utilizing contrast limited adaptive histogram equalization (CLAHE) that decreases the noise hindrance in the images and shadow of images. CLAHE method has been utilized for enhancing the degraded contrast of the mammogram images. It reduces the shadow of the edge along with produces noise in the images [5]. It is calculated in our proposed equation (1).

$$CE = p_t(A_f) = \frac{n_f}{n} \quad (1)$$

where *CE is* contrast enhancement, $p_t(A_f)$ is probability density function of the input patch, f is input pixel, m is total number of pixels, $n_f$ is gray scale value of input and t is clip limit.

Equation (1) has been modified to form a modified contrast enhancement as equation (2). This transformation function is used to change the image RGB value.

$$MCE = \sum_{f=0}^{t} p_t(A_f) \quad (2)$$

where MCE is modified contrast enhancement, $p_t(A_f)$ and t are same as in (1), and f is RGB value of input pixel

The optical density (OD) with a logarithmic function in state of the art is modified by adding a modified contrast enhancement from equation (2) into equation (3). This function normalizes the concentration of the stained images along with a reduction in the noise and edges [5]

$$MOD = OD + MCE \quad (3)$$

Finally, the augmented patch dataset that is calculated in the state of the art is modified by us to be equation (4). Now, this new equation helps in enhancing the contrast and quality of the input breast images. It also reduces the noise amplification that is caused by those input images.

$$MPa_i(a,b) = a_d(p_d(MOD)) \quad (4)$$

where MOD is modified optical density, CE is colour enhancement function, $A_d$ is augmented data and $P_d$ is patched data

Equation (5) shows the ReLU function which filters out all the negative information. ReLU causes a small derivative and is expressed as in [1]. It solves the problem of vanishing gradient error prevalent when using Sigmoid and tangent functions.

$$r(z) = \begin{cases} 0 & \text{for } z < 0 \\ z & \text{for } z \geq 0 \end{cases} \quad (5)$$

where z is input and r is ReLU function

Similarly, to reduce the gradient error minimizing the negative points, ArcTan is used [1] and presented in equation (6) as:

$$t(z) = \tan^{-1}(z) \quad (6)$$

where t is ArcTan function and z is same as in (5)

For the proposed solution a modified Tanh function is used with the combination of ReLU and ArcTan and it is expressed as in equation (7):

$$Mr(z) = \begin{cases} \tan^{-1}(z) & \text{for } z < 0 \\ z & \text{for } z \geq 0 \end{cases} \quad (7)$$

Algorithm for reducing input dimension using convolutional neural network in table 1, step 6 is replaced by equation (7) as shown in table 2.

Table 2: updated Convolutional neural network with modified tangent function algorithm

| |
|---|
| **Algorithm2: Updated Convolutional method and max-pooling method of convolutional neural network for reducing the input dimension.** |
| Step7: Apply the modified tangent function to step 5 to obtain energy output<br>$Mr(z) = \begin{cases} \tan^{-1}(z) & \text{for } z < 0 \\ z & \text{for } z \geq 0 \end{cases}$<br>Where z = input |
| END |

We modified equation (3) to be equation (8) which shows the modified output feature map. This output feature map uses the modified tangent activation function. It reduces the vanishing gradient error by minimizing and filtering the negative points and causing a small derivative.

$$P'_{(a,b)} = Mr(z) * \left( \sum_{x=-u}^{u} \sum_{m=-y}^{y} C(x,m) \, Pa_i (a-m, b-x) \right) \quad (8)$$

Therefore, equation (8) is enhanced by equation (4) to bring our final enhanced proposed formula in equation (9). This final equation is used to extract discriminating features from the image and generates an output feature map by the user for a modified tangent activation function and image enhancement function.

$$EP'_{(a,b)} = Mr(z) * \left( \sum_{x=-u}^{u} \sum_{m=-y}^{y} C(x,m) \, Pa_i (a,b) \, (a-m, b-x) \right) \quad (9)$$

Enhancing the image quality and reducing the noise improves the visual contrast of the images. When images are visualized correctly, there are more chances to predict accurate cancer types. Similarly, using a modified tangent activation function aim to reduce the vanishing gradient error. During the backpropagation, the gradient of loss is calculated according to the weights. The gradient tends to grow smaller as going backward in the neural network. It means that neurons in the earlier layer learn slowly in comparison to the neuron in the later layers in a hierarchy. The earlier layer is slower to train in the network. The earlier layer is the building block of the neural network. So, if they provide an improper and inaccurate result, then the output provided by later layers also produces an inaccurate result in further processing. Thus, the reduction in the vanishing gradient error automatically increases the accuracy of the system for predicting cancer. As seen from the literature presented in this paper, the existing solutions that applied convolutional neural networks have only used color normalization techniques and hyperbolic tangent function. It hasn't considered noise, better contrast of input images, and gradient error. Our proposed solution has enhanced the contrast of the image and reduces the noise produced along with the reduction of vanishing gradient error. It increases the accuracy of the classification of breast cancer images by using the CLAHE method and modified tangent function.

## 4. Results and Discussion

Python 2.7.1 was utilized in the implementation and simulation of a model using four types of sample images from various stages of a cancerous cell in [3]. For implementation, we used the system configuration as: CPU based system with Intel Xeon Processor, 128 GB RAM. Python 3.1.1 is used with Tensorflow and Keras library. The datasets were collected from the ICIAR 2018 challenge. For the experiment, a 2.2 GHz Intel Core i7 processor with 16 GB RAM is used. More than 400 different stained breast histology images are used for training purposes. The training datasets were distributed in each class and it is annotated into 4 different stages namely normal, benign, in-situ, and invasive carcinoma. However, the class label of a hidden test dataset was not disclosed. As a practice, the data set was divided into three parts for training, validation, and testing. 10 images from each histological class in the training dataset were separated from reporting the image-wise classification performance of our model. 10% of 34,560 samples that is 3456 samples are used for the validation purpose. Cross-validation is used where the value of k is 10. Sample data is randomly divided into k equal-sized subsample where one is used for testing and the remaining others for the training. The validation measure is calculated by classification accuracy. Accuracy is calculated by the sum of true negative and true positive to the sum of a false positive, false negative, true positive, and true negative. The performances of the proposed solution with the state of the art solution are compared based on accuracy percentage time which is shown in table 3 and table 4 respectively. Finally, the results in terms of processing time are obtained as an average for all the tested cases.

Accuracy is the most commonly used assessment measure for the classification purpose that is considered all the cases; it used all the cases. See equation 10 [3].

$$\text{Accuracy} = \frac{TN + TP}{FP + FN + TP + TN} \qquad (10)$$

where     TN is True Positive
               TP is True Positive
               FP is False Positive
               FN is False Negative

During the pre-processing stage, the images are passed through the CLAHE algorithm to increase the contrast of the image along with noise reduction. See Fig. 3. Data augmentation is performed next.

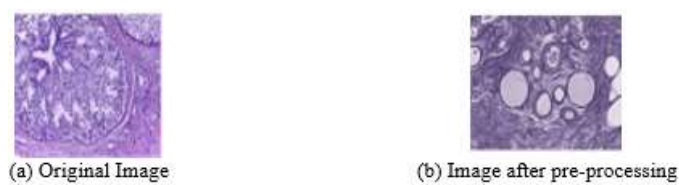

Fig 3 Contrast Limited Adaptive Histogram Equalization (CLAHE)

After pre-processing the input image is passed through a convolutional neural network for extracting the features of the image through different layers. After feature extraction, the image is passed to the SoftMax layer that contains the output neuron for classification. Further, a patch-based classifier is used to predict the class label as normal, benign, in-situ, and invasive carcinoma. Patch-wise classification is done first. Patch labels of an image being predicted by OPOD technique and, the result of this classification is again passed for image-wise classification. During this process, the majority voting scheme is done to get better accuracy in the classification process.

Experiments were conducted using different datasets namely, ICIAR 2015, MIAS, ICIAR 2016, BreakHis, ICIAR 2017, and ICIAR 2018 that contains the images of normal, malignant, in-situ, and invasive carcinoma cells. The results from the breast cancer samples are presented in **Tables 3, 4,5, 6, 7 and 8. Each of these tables show the** result comparison between the state of the art and the proposed solutions with the help of graphs and the data reports. The results are divided according to the four kinds of cancer cells to see the impact of noise on it. Here, the results from the sample are presented in terms of accuracy and processing time. Accuracy is calculated in terms of true positive, false positive, true negative, and false negative. We have performed comprehensive evaluation using 15 tests; 3 tests in each scenario and each test has 4 cases; normal, benign, in-situ, and invasive carcinoma. The accuracy result has been calculated by taking the average result of each test case. Then the final result has been calculated by taking the average of all the tested cases in three scenarios. These results were compared during different image processing stages. The results showed that the proposed solution has improved the accuracy of classification as normal, benign, in-situ, and invasive carcinoma.

Table 3: Accuracy and Processing time for classification of breast cancer (Sample image group 1)

| Sample No. | Sample group details | Original Images | State of the Art | | | Proposed solution | | |
|---|---|---|---|---|---|---|---|---|
| | | | Processed sample | Accuracy (%) | Processing time (sec) | Processed sample | Accuracy (%) | Processing time (sec) |
| 1.1 | ICIAR 2015 | 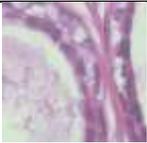 | 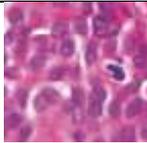 | 89% | 0.55s | 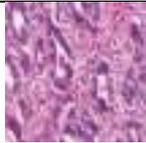 | 92% | 0.42s |
| 1.2 | | 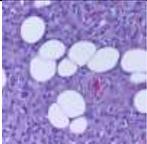 | 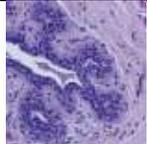 | 86% | 0.40s | 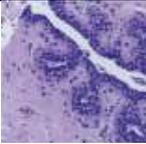 | 90.45% | 0.39s |
| 1.3 | | 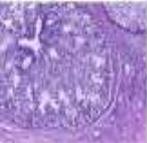 | 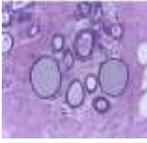 | 86.16% | 0.412s | 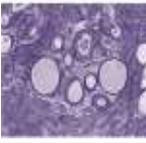 | 90.77% | 0.37s |
| 1.4 | | 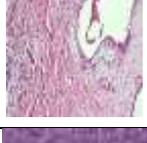 | 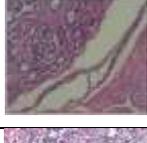 | 85.78% | 0.39s | 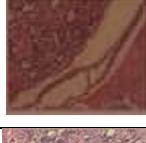 | 91.67% | 0.31s |
| 1.5 | | 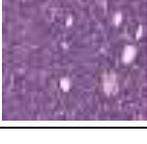 | 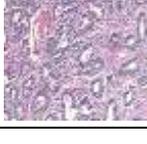 | 83.34% | 0.43s | 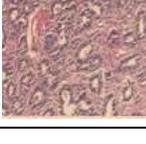 | 89% | 0.33s |

Table 4: Accuracy and Processing time for classification of breast cancer (Sample image group 2). The green border shows the cancer tissues present in the breast.

| Sample No. | Sample group details | Original Images | State of the Art | | | Proposed solution | | |
|---|---|---|---|---|---|---|---|---|
| | | | Processed sample | Accuracy (%) | Processing time (sec) | Processed sample | Accuracy (%) | Processing time (sec) |
| 2.1 | | 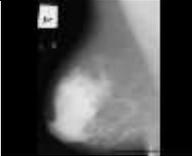 | 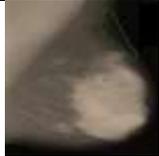 | 87.01% | 0.409s | 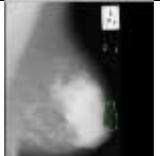 | 94.63% | 0.31s |
| 2.2 | | 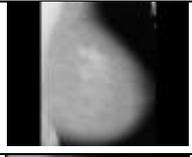 | 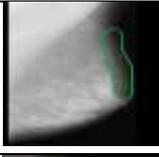 | 86.16% | 0.373s | 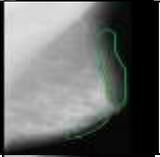 | 92.67% | 0.28s |
| 2.3 | **MIAS dataset** | 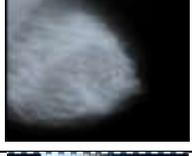 | 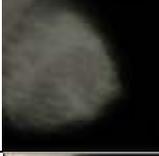 | 86.03% | 0.384s | 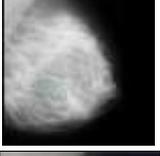 | 91.62% | 0.341s |
| 2.4 | | 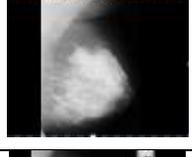 | 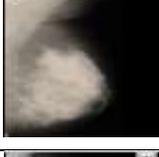 | 87.34% | 0.456s | 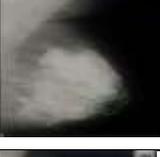 | 91.89% | 0.24s |
| 2.5 | | 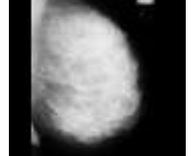 | 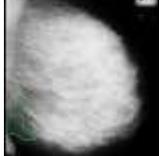 | 87% | 0.43s | 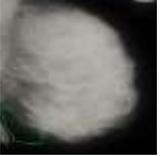 | 94% | 0.21s |

Table 5: Accuracy and Processing time for classification of breast cancer (Sample image group 3)

| Sample No. | Sample group details | Original Images | State of the Art | | | Proposed solution | | |
|---|---|---|---|---|---|---|---|---|
| | | | Processed sample | Accuracy (%) | Processing time (sec) | Processed sample | Accuracy (%) | Processing time (sec) |
| 1.1 | Images of breast histopathological images ICIAR 2016 | 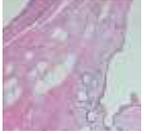 | 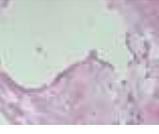 | 90% | 0.31s | 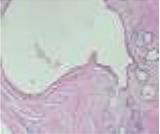 | 92.05% | 0.30s |
| 1.2 | | 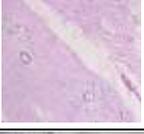 | 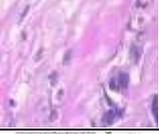 | 88.22% | 0.35s | 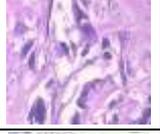 | 93.35% | 0.31s |
| 1.3 | | 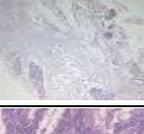 | 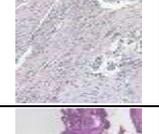 | 91.42% | 0.44s | 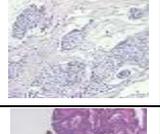 | 94.00% | 0.31s |
| 1.4 | | 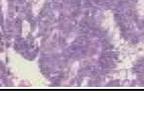 | 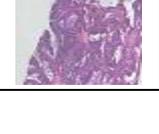 | 84.65% | 0.45s | 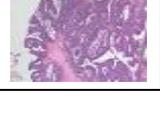 | 93.90% | 0.30s |

Table 6: Accuracy and Processing time for classification of breast cancer (Sample image group 4)

| Sample No. | Sample group details | Original Images | State of the Art | | | Proposed solution | | |
|---|---|---|---|---|---|---|---|---|
| | | | Processed sample | Accuracy (%) | Processing time (sec) | Processed sample | Accuracy (%) | Processing time (sec) |
| 1.2 Ductal Carcinoma | Images of breast histopathological images BreakHisv_1 | 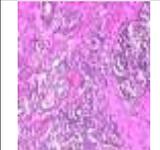 | 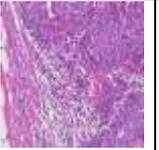 | 82% | 0.35s | 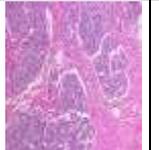 | 86.05% | 0.27s |
| 1.2 Lobular carcinoma | | 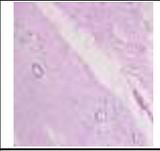 | 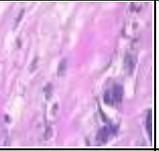 | 89.22% | 0.47s | 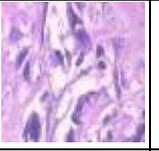 | 90.35% | 0.39s |
| 1.3 Mucinous carcinoma | | 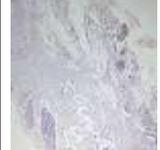 | 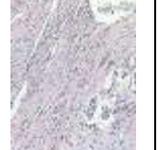 | 85.56% | 0.54s | 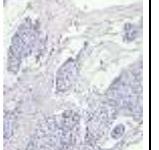 | 93.78% | 0.41s |
| 1.4 Papillary carcinoma | | 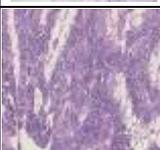 | 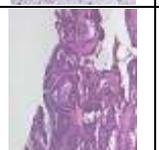 | 84.65% | 0.39s | 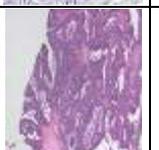 | 88.90% | 0.27s |

Table 7: Accuracy and Processing time for classification of breast cancer (Sample image group 5)

| Sample No. | Sample group details | Original Images | State of the Art | | | Proposed solution | | |
|---|---|---|---|---|---|---|---|---|
| | | | Processed sample | Accuracy (%) | Processing time (sec) | Processed sample | Accuracy (%) | Processing time (sec) |
| 1.2 Adenosis | Images of breast histopathological images ICIAR 2018 | 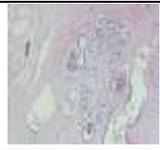 | 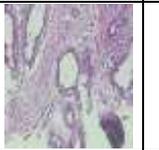 | 91% | 0.21s | 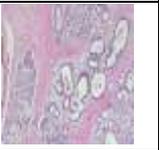 | 90.05% | 0.21s |
| 1.2 Fibroadenoma | | 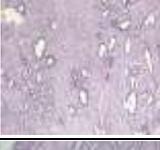 | 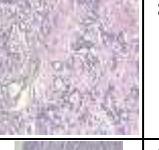 | 87.22% | 0.33s | 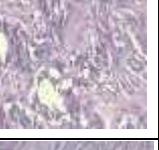 | 88.35% | 0.31s |
| 1.3 Phyllodes tumor | | 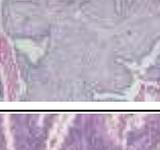 | 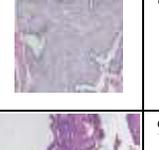 | 82.42% | 0.44s | 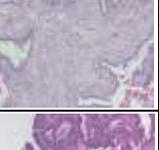 | 91.78% | 0.31s |
| 1.4 Papillary carcinoma | | 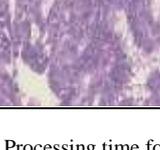 | 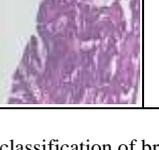 | 90.63% | 0.33s | 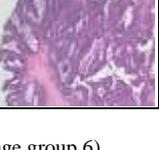 | 93.10% | 0.30s |

Table 8: Accuracy and Processing time for classification of breast cancer (Sample image group 6)

| Sample No. | Sample group details | Original Images | State of the Art | | | Proposed solution | | |
|---|---|---|---|---|---|---|---|---|
| | | | Processed sample | Accuracy (%) | Processing time (sec) | Processed sample | Accuracy (%) | Processing time (sec) |
| 1.2 Ductal Carcinoma | Images of breast histopathological images ICIAR 2017 | | | 87% | 0.31s | | 89.05% | 0.30s |
| 1.2 Lobular carcinoma | | | | 87.22% | 0.45s | | 90.35% | 0.32s |
| 1.3 Mucinous carcinoma | | | | 89.22% | 0.64s | | 90.78% | 0.41s |
| 1.4 Papillary carcinoma | | | | 83.55% | 0.40s | | 88.40% | 0.37s |

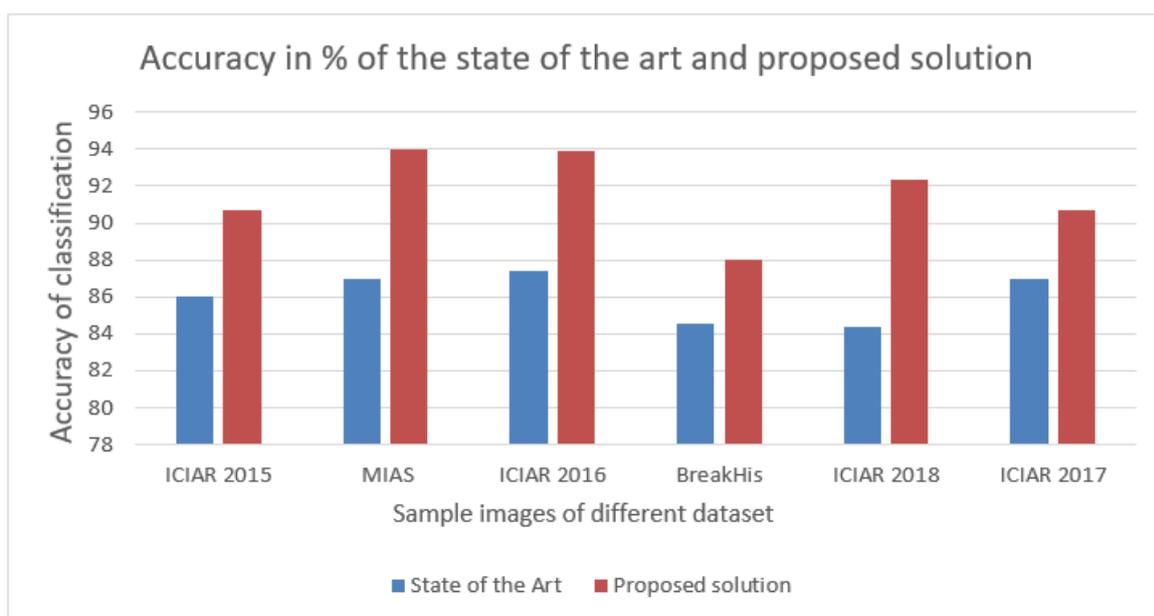

Fig. 4: Bar graph shows the classification accuracy in percentage for six different datasets of histopathological images.

Figure 4 shows accuracy presented in percentage comparison of base (state of the art) system and propsed system for datasets ICIAR 2015, MIAS, ICIAR 2016, BreakHis, ICIAR 2018 and ICIAR 2017. The blue color indicates the accuracy of the state of the art solution whereas the orange colour indicates the accuracy of the Proposed Solution. The first, third, fifth, and sixth couple of bar graphs show the average accuracy for ICIAR 2015, 2016, 2018, and 2017 dataset respectively. (2) A second couple of bar graphs shows the average accuracy for the MIAS dataset. (3) A fourth couple of bar graphs shows the average accuracy for the BreakHis dataset.

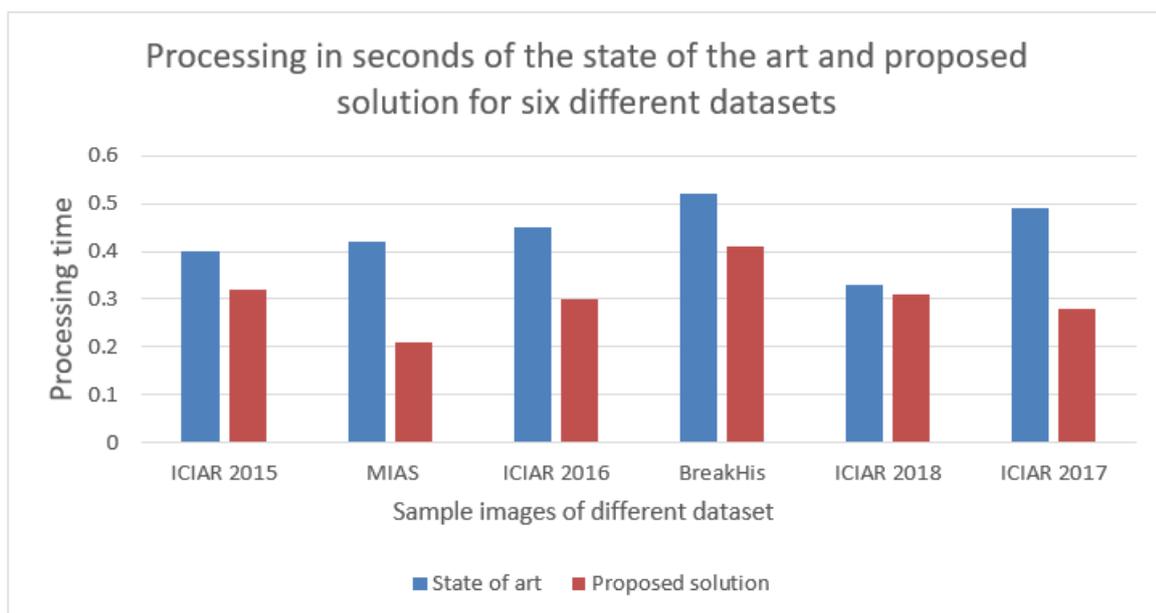

Fig. 5: The bar graph shows the processing time in seconds for six different datasets of histopathological images.

Figure 5 shows processing time comparison of base (state of the art) system and propsed system for datasets, ICIAR 2015, MIAS, ICIAR 2016, BreakHis, ICIAR 2018 and ICIAR 2017. The blue color indicates the accuracy of the state of the art solution whereas the orange color indicates the accuracy of the proposed Solution. First, third, fifth, and sixth couple of bar graph shows the average accuracy for ICIAR 2015, 2016, 2018 and 2017 dataset respectively. (2) A second couple of bar graphs shows the average accuracy for the MIAS dataset. (3) A fourth couple of bar graphs shows the average accuracy for the BreakHis dataset.

Results showed the difference in accuracy and processing time between the current state of the art solution and the proposed solution concerning the classified image sets. The proposed algorithm works on improving the classification accuracy of histopathology images to 94% with the help of modified tangent function and (CLAHE) technique, which is considered better in comparison to use of tanh function in the state of the art solution. Similarly, the processing time for the classification process decreased between 0.2~ 0.4 seconds. The obtained result is better than the result of the state of the art solution. The accuracy and processing time are calculated by simulating the temporal protocol in Python systems. Accuracy is calculated by finding the sum of true positive and true negatives over the sum of a false positive, false negative, true negative, and false negative. We quantify the degree of improvement in processing time by running the state of the art and proposed algorithms and the duration of running each algorithm. Accuracy is measured in terms of percentage, and the processing time is calculated in seconds.

The current convolutional network uses a modified tangent function along with a convolutional layer, max-pooling layer, drop out, and SoftMax layer. The modified tangent function is designed by combining two other functions namely ReLU and arcTan function. ReLU and arcTan together to backpropagate the errors and to activate the multiple layers of a neuron. This lowers the run time i.e. processing time is decreased while increasing the accuracy rate for classification. Moreover, the solution uses an image enhancement algorithm called (CLAHE) for enhancing the contrast of an image and reducing the noise amplification. Due to this, an image with detailed information can be obtained which is very crucial for classifying accurately and faster. To conclude, the convolutional neural network combined with a patch-based classifier has provided improved accuracy by 6% and processing time by 0.2~0.4 seconds.

Some different algorithms and techniques have been used for classifying the breast images. However, number of alternatives have been reviewed to improve the accuracy and processing time of the classification. The limitation of the current state of the art system has been solved with an improved rate of accuracy i.e. 94% against the current accuracy of 87%. The proposed system has reduced the processing time from 0.56 seconds to 0.23 seconds. Such improvement in accuracy and processing time is possible due to the addition of (CLAHE) and modified tangent function solving the gradient error problem. Henceforth, the proposed system has improved accuracy and reduced processing time in all the taken samples. A comprehensive comparison between state of the art and proposed solution is done in Table 9.

Table 9. Comparison table between Proposed Solution and State of the Art Solution

|  | Proposed solution | State of the Art Solution |
|---|---|---|
| **Name of the solution** | Modified color equalization method and Contrast Limited Adaptive Histogram Equalization (CLAHE) | Enhanced color normalization technique with an optical density |
| **Image Enhancement** | Improve the contrast of the image in terms of shadow edges and noise produced. It provides a significant contrasting image with image improvement. | The enhanced image in terms of contrast. |
| **Function Name** | Modified hyperbolic tanh activation function (ReLU and ArcTan) | Uses hyperbolic tanh activation function. |
| **Accuracy** | Improved accuracy in terms of vanishing gradient error. | Provides an accuracy of 87%. |
| **Processing Time** | The decrease in processing time from 0.56 seconds to 0.24 second | Provides a processing time of 0.24 seconds. |
| **Proposed equation** | $P'_{(a,b)} = \text{Mr}(z) * (\sum_{x=-u}^{u} \sum_{m=-y}^{y} C(x,m) Pa_i (a-m, b-x))$ | $P_{(a,b)} = g(z) * (\sum_{x=-u}^{u} \sum_{m=-y}^{y} C(x,m) Pa_i (a-m, b-x))$ |
| **Contribution 1** | Contrast Limited Adaptive Histogram Equalization (CLAHE) method enhances the degraded contrast of the images by transforming the pixel intensity to the value within a display range proportional to a pixel intensity's rank in a local intensity histogram. It uses the **clip limit** that helps maximize the contrast enhancement. It is introduced to reduce the noise produced and shadow of edges in the breast cancer images. | The state of the art system does not include the problem created by the degraded contrast and noise produced by the images in terms of accuracy. |
| **Contribution 2** | Modified tangent function solves the problem of vanishing gradient error that occurs during the backpropagation. It helps in filtering out all negative information | The state of the art system does not reduce the vanishing gradient error produced by the use of a hyperbolic tangent function. |

# 5 . Conclusions and Future Work

**The proposed solution with differential information about the contrast and noise reduction helped in increasing the accuracy of the classification.** This is a new feature adapted from the second-best solution. This information increases the contrast of input breast histology images and reduces the noise produced while processing the images. This maximizes the image quality, which helps the system to take clear images for training and testing. Finally, the accuracy of classifying a type of cancer in the breast will be increased. With this experiment, the idea of the second base solution has proven to increase the accuracy by 5% by using the (CLAHE) technique. Similarly, using a modified tangent hyperbolic activation function in a feature extraction process with the use of a convolutional neural network decreases the vanishing gradient problem along with over-fitting issues. **Also, a combination of convolutional neural networks with a patch-based classifier provided improved accuracy by 6% and processing time by 0.2~0.4 seconds.** The proposed system has been tested using Python simulator and the results have shown the accuracy is increased of the overall classification. **The negative points are decreased by using ReLU instead of eLU, which is the eLU can produce negative outputs. In future work, a combination of a** single image-based classification with the addition of embedded technology such as SoftMax classification **would be explored**. Furthermore, if patch-based classification is removed then ultimately patch extraction process can be removed from the pre-processing stage. This removal can decrease the processing time further.

# 6 . Appropriate Data Availability Statement

No real data have used in this work. All data sets are freely available in Google and have downloaded by us